\documentclass[aip,sd,amsmath,amssymb,reprint]{revtex4-1}

\usepackage{graphicx}
\usepackage{dcolumn}
\usepackage{bm}
\usepackage{color}
\usepackage{adjustbox}

\begin{document}

\title{Control, bi-stability and preference for chaos in time-dependent vaccination campaign}

\author{Enrique C. Gabrick} 
\email{ecgabrick@gmail.com}
\affiliation{Potsdam Institute for Climate Impact Research, Telegrafenberg A31, 14473 Potsdam, Germany.}
\affiliation{Department of Physics, Humboldt University Berlin, Newtonstra{\ss}e 15, 12489 Berlin, Germany.}
\affiliation{Graduate Program in Science, State University of Ponta Grossa, 84030-900, Ponta Grossa, PR, Brazil.}
\author{Eduardo L. Brugnago}
\affiliation{Institute of Physics, University of S\~ao Paulo, 05508-090, S\~ao Paulo, SP, Brazil.}
\author{Ana L. R. de Moraes}
\affiliation{Department of Physics, State University of Ponta Grossa, 84030-900, Ponta Grossa, PR, Brazil.}
\author{Paulo R. Protachevicz}
\affiliation{Institute of Physics, University of S\~ao Paulo, 05508-090, S\~ao Paulo, SP, Brazil.}
\author{Sidney T. da Silva} 
\affiliation{Department of Chemistry, Federal University of Paraná, 81531-980, Curitiba, PR, Brazil.}
\author{Fernando S. Borges}
\affiliation{Graduate Program in Science, State University of Ponta Grossa, 84030-900, Ponta Grossa, PR, Brazil.}
\affiliation{Department of Physiology and Pharmacology, State University of New York Downstate Health Sciences University, 11203, Brooklyn, New York, USA.}
\affiliation{Center for Mathematics, Computation, and Cognition, Federal University of ABC, 09606-045  S\~ao Bernardo do Campo, SP, Brazil.}
\author{Iber\^e L. Caldas}
\affiliation{Institute of Physics, University of S\~ao Paulo, 05508-090, S\~ao Paulo, SP, Brazil.}
\author{Antonio M. Batista}
\affiliation{Graduate Program in Science, State University of Ponta Grossa, 84030-900, Ponta Grossa, PR, Brazil.}
\affiliation{Institute of Physics, University of S\~ao Paulo, 05508-090, S\~ao Paulo, SP, Brazil.}
\affiliation{Department of Mathematics and Statistics, State University of Ponta Grossa, 84030-900, Ponta Grossa, PR, Brazil.}
\author{J\"urgen Kurths}
\affiliation{Potsdam Institute for Climate Impact Research, Telegrafenberg A31, 14473 Potsdam, Germany.}
\affiliation{Department of Physics, Humboldt University Berlin, Newtonstra{\ss}e 15, 12489 Berlin, Germany.}

\begin{abstract}
In this work, effects of constant and time-dependent  vaccination
rates on the Susceptible-Exposed-Infected-Recovered-Susceptible (SEIRS) seasonal model are studied.
Computing the Lyapunov exponent, we show that typical complex structures,
such as shrimps, emerge for given combinations of constant vaccination rate and 
another model parameter. In some specific cases, the constant vaccination 
does not act as a chaotic suppressor and
chaotic bands can exist for  high levels of vaccination (e.g., $> 0.95$). Moreover,
we obtain linear and non-linear relationships between one control parameter
and constant vaccination to establish a disease-free solution. We also verify that the
total infected number does not change whether the dynamics is chaotic or periodic.   
The introduction of a time-dependent vaccine is made by the inclusion of a periodic function
with a defined amplitude and frequency. For this case,
we investigate the effects of different amplitudes and frequencies on chaotic attractors,  
yielding low, medium, and high seasonality degrees of contacts.
Depending on the parameters of the time-dependent vaccination function, chaotic
structures can be controlled and become periodic structures.
For a given set of parameters, these structures are accessed mostly via crisis
and in some cases via period-doubling. After that, we investigate how
the time-dependent vaccine acts in
bi-stable dynamics when chaotic and periodic attractors coexist.
We identify that this kind of vaccination acts as a control by destroying
almost all the periodic basins. We explain
this by the fact that chaotic attractors exhibit more desirable  
characteristics for epidemics than periodic ones in a bi-stable state.
\end{abstract}

\maketitle

\begin{quotation}
Mathematical models are a powerful 
tool to study the spread of diseases and strategies to control them.
In the present contribution, we examine the effects of constant and 
time-dependent vaccination campaigns on the dynamical behaviour of a
seasonal forced Susceptible-Exposed-Infected-Recovered-Susceptible (SEIRS) 
model. The impacts of such strategies are investigated through Lyapunov 
exponents for parameter planes.
We discover that 
typical complex structures, 
emerge for constant vaccination and for a certain parametric configuration. 
 The structures 
observed in the parameters planes show very complex dynamics in the model. 
Next, we use parameters for chaotic and bi-stable solutions and explore the 
influence of a time-dependent immunisation program. 
 We uncover that the time-dependent 
vaccination campaign can control the chaotic bands, making 
them periodic under specific conditions. 
Using a parametric configuration that generates bi-stable solutions 
where periodic-periodic and chaotic-periodic attractors coexist, we 
investigate the effects of a time-dependent immunisation campaign 
on the dynamics. For this,   
we develop a method to select  
one attractor over another. Additionally, our results show that 
chaotic attractors spend more time at low infection levels than periodic 
ones for the same parametric configuration.   
\end{quotation}


\section{Introduction}
Vaccination is one of the most efficient ways to control disease  
spreading \cite{Keeling2008}. For example, in United Kingdom (UK) the reported annual 
cases of measles in the pre-vaccination  were around 100,000 to 800,000, 
but after the introduction of vaccination in 1968 the reported cases 
decreased to 30,000 per year, from 1968 up to 1988 \cite{Jick2010}. More recently, it was estimated that 
COVID-19 vaccination prevented around 14 million deaths in 185 countries, 
from Dec. 2020 to Dec. 2021 \cite{Watson2022}. 

Vaccine campaigns can be conducted by different strategies. One is the mass 
vaccination campaign which  consists of vaccinating  
a large number of people 
in a short time \cite{Grabenstein2006}. 
Another is to vaccinate different groups in discrete time intervals, namely 
pulsed vaccination \cite{Agur1993}.  
Moreover, some diseases are seasonal and are better controlled by combining 
seasonal vaccination with routine campaigns \cite{Alvarez2009}, i.e., make them 
time-dependent. 
Such strategies have been shown useful against certain diseases \cite{Williams1983}, 
especially seasonal ones \cite{Greenwood2017}. When they are combined 
with routine  programs, the time-dependent term causes perturbations 
that can drive the system to a new equilibrium situation \cite{Coutinho1993}. 

One form to decide the vaccination strategy is by means of
 mathematical models \cite{Massad1995}, 
once these tools are powerful for modelling and 
studying mitigation strategies for epidemic spread \cite{AndersonMay}. 
In this way, mathematical models bring very important insights  
for the elaboration of campaigns. In a situation where two vaccine doses need to be administrated, the models 
suggest that is better to descriminate the population before applying the vaccine \cite{Gabrick2022}. 
With this procedure, a smaller amount of doses is needed to reach the same effects as when 
they are applied randomly. In addition, if the vaccine is applied in a 
pulsed protocol, the number of wasted doses is drastically reduced.
 When seasonal 
effects are taken into account in the contact rates, pulsed vaccine combined  
with routine is more efficient than just routine vaccine \cite{Gabrick2024}. 
Many works have been done in terms of vaccination strategies 
\cite{Liu2008, Wang2019, Metcalf2012, Thater2018, Biswas2014, Gao2008b}. 
However, there are only few works dedicated to understand the
 dynamical behaviour of such systems. 

Under constant vaccination protocols, the SIR model with logistic growth can produce 
complex dynamics for a set of parameters \cite{Carvalho2023}. These solutions 
exhibit different kinds of bifurcation such as Hopf, transcritical, Belyakov 
and saddle-node, depending on the basic reproduction number ($R_0$) and 
the proportion of vaccinated individuals. 
When pulses are incorporated into the vaccination strategy, solutions of 
disease-free (DFE) can be obtained depending on the rates \cite{Shulgin1998}. 
Moreover, the system is driven to chaotic solutions in the presence of 
seasonal variations. 
The chaotic solutions generated due to the seasonal forcing can be controlled  
to periodic motion when a seasonal component is added to the vaccination 
strategy \cite{Duarte2021}. 
For a general periodic vaccination rate, periodic DFE solutions are ensured when $R_0 < 1$ 
and have  the same period as the vaccine function \cite{Moneim2005}. 
DFE solutions are globally asymptotically stable for $R_0 < 1$ in a situation 
where the period of vaccination function is a multiple integer of the 
seasonal contact rate. Nonetheless, this solution becomes unstable for $R_0 > 1$ \cite{Moneim2005b}. 
In a SEIRS seasonal model, chaotic and bi-stable solutions can emerge 
for constant and pulsed vaccination rates \cite{Gabrick2023}.

{In this work, we study the dynamical behaviour of a seasonal forced SEIRS   
model under constant and non-constant (seasonal) vaccination rates. 
We explore the dynamical properties in a wide range of parameters and present 
a control method for bi-stability. In this way, this research is a 
 strong generalization of a previous paper about the effect of the constant 
and pulsed protocol of vaccination rates \cite{Gabrick2023}.} 
We investigate the effects of constant and time-dependent 
vaccination rate into the dynamical behaviour of a seasonal SEIRS model, 
aiming to understand the following questions: ($i$) Could the constant vaccination rate 
 suppress chaotic dynamics? ($ii$) What are the effects of vaccination types   
in the bi-stable range?

To explore these questions, we split this work into two parts.
In the first one, for a constant vaccination protocol, 
we verify that chaotic solutions  
can be produced even for high vaccination rates depending on 
the other parameters. 
Furthermore, we show that under specific conditions some typical complex 
structures such as shrimps \cite{Gallas1993} emerge. This typical complex 
structures {were firstly named by J. A. C. Gallas \cite{Gallas1993}}. 
In terms of bi-stability, we find that 
the seasonal vaccination can act as a control method.
This control enables us to 
select the attractors as a function of the vaccination parameters. 
In addition, periodic basins can be destroyed
due to high levels of infection.

We organise this work as following: In Section \ref{sec_model}, we present the model 
and discuss some aspects of its equilibrium. Section \ref{sec_constant} is 
dedicated to show the results related to constant vaccination. The seasonal 
vaccination rate is described in Section \ref{sec_seasonal} and its influence 
on bi-stable solutions is discussed in Section \ref{section_bistability}. 
Finally, we draw our conclusions in Section \ref{sec_conclusion}.

\section{Model}\label{sec_model}
\begin{figure}[!ht]
	\centering
	\includegraphics[scale=0.3]{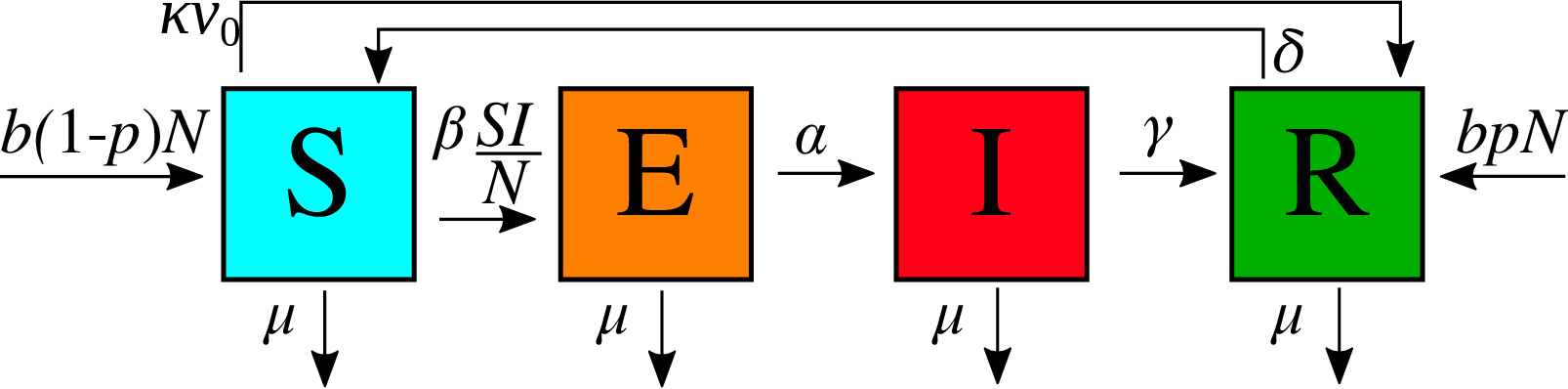}
	\caption{Representation of SEIRS model with vaccination.} 
	\label{fig1}
\end{figure}

Given a population of size $N$, the SEIRS model splits the population 
according to their infectious status, which are 
Susceptible ($S$), Exposed ($E$), Infected ($I$), and Recovered ($R$) \cite{Bjornstad2018}. 
Healthy individuals are stored in $S$, infected but not infectious in $E$, 
infectious in $I$, and recovered in $R$ \cite{Keeling2008}. Originally, 
this model did not include vaccination \cite{Bjornstad2020}. However, this can 
be done by transferring newborns and other $S$ individuals   
to $R$ compartment \cite{Bai2012}, as schematically represented in Fig. \ref{fig1}. 
The arrows indicate the flow between the compartments, where the parameters are: 
$b$ birth rate; 
$\beta$ contact rate of between $S$ and $I$ individuals; 
$1/\alpha$ latent period; 
$\gamma$ recovery rate; 
$1/\delta$ time to loss the immunity; 
$\mu$ natural death rate; 
$\kappa$ vaccine efficacy; 
{$p$ newborns vaccination rate;}
and $v_0$ rate at which $S$ individuals are vaccinated. Without loss 
of generality, we use $\kappa=1$ in this whole work. Observe that the 
fraction $bpN$ that is introduced in $R$ came from the vaccination in 
the $S$ newborns.

The representation in Fig. \ref{fig1} shows the compartments $S$, $E$, $I$, 
and $R$ in capital letters, which represent the number 
of individuals in each compartment, from a given population $N$. From the 
modelling point of view, we can work with the model independent of $N$. 
To do that, we write the equations in terms of the fractions $s$, $e$, $i$, 
and $r$. A complete discussion about the normalization is given in Ref. \cite{Brugnago2023}. 
The normalised SEIRS is described by the following system of ordinary differential equations 
\begin{eqnarray} 
		\frac{ds}{dt} & =& b(1-p) - \beta{s i} + \delta (1-s-e-i) \nonumber \\
		 &-& \mu s - \kappa v_0 s, \label{n1}\\
		\frac{de}{dt} & =& \beta {s i} - (\alpha + \mu)e,  \label{n2}\\
		\frac{di}{dt} & =& \alpha e - (\gamma + \mu)i,  \label{n3}
\end{eqnarray}
where we assume $b=\mu$, then, $r=1 - s - e - i$. For more details about 
the equilibrium solutions and properties of Eqs. (\ref{n1})-(\ref{n3}) 
refer to \cite{Gabrick2024}. 

We consider a time-dependent contact rate given by 
\begin{equation}
	\beta \equiv \beta(t) = \beta_0 \left[1 + \beta_1 {\rm cos} (\omega t)\right],
	\label{eq_betaFuncao}
\end{equation} 
where $\beta_0$ is the average contagion rate, $\beta_1\in [0,1]$ is the 
seasonality degree, and $\omega$ is the frequency. This formulation 
is taken to model seasonal infectious diseases transmission \cite{Altizer2006,Earn2000b,Grassly2006}.

The main novelty of this work is to explore  effects of a seasonal 
vaccination, that is modelled by the continuous
 function: 
\begin{equation}
v(t) = v_0 + \xi {\rm cos} (\omega_{\rm v} t),
\label{vaccinaperturbada}
\end{equation}
where $\xi \in [0, v_0 ]$ is the amplitude and $\omega_{\rm v}$ is 
the frequency of the vaccination campaign. 
Observe that when $\xi \ll 1$ the vaccination campaign is predominantly 
given by $v_0$ but with a periodic perturbation. 
The period of vaccination is given by 
$T_{\rm v} = 2\pi/\omega_{\rm v}$ in years unity.

The complete model is described by Eqs. (\ref{n1})-(\ref{n3}) 
together to Eq. (\ref{eq_betaFuncao}) and (\ref{vaccinaperturbada}). 
For biological reasons, the initial
conditions and the parameters present in Eqs. (\ref{n1})-(\ref{n3}), (\ref{eq_betaFuncao}) and (\ref{vaccinaperturbada}) 
are $\geq 0$. Then, the solutions are bounded in the set: $D = \{(s,e,i) \in [0,1]^3 \}$ 
for every $t\geq0$ (Ref. \cite{Gao2008}). 
The solutions are numerically obtained by 4th order Runge-Kutta method \cite{Boyce2012} 
with the integration step equal to $10^{-3}$. 

\begin{figure}[!ht]
	\centering
	\includegraphics[scale=0.8]{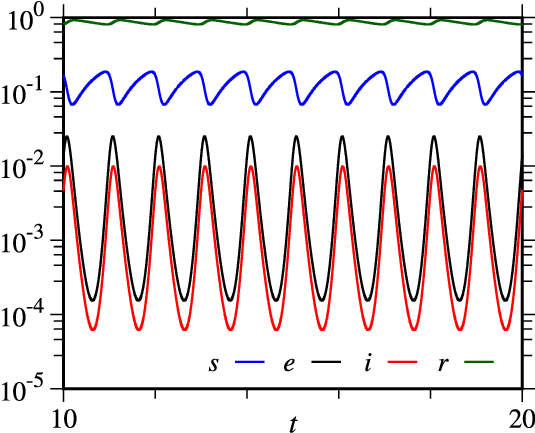}
	\caption{Numerical solution for SEIRS model. The blue, black, red, and green 
	lines are for $s$, $e$, $i$, and $r$, respectively. We consider 
	$b = 0.02$, $p = 0.25$, $\kappa = 1$, $v_0 = 0.2$, $\beta_{0} = 800$, 
	$\beta_{1} = 0.20$, $\omega = 2\pi$, $\alpha = 40$, $\gamma = 100$, 
	$\delta = 0.25$, {and $\xi = 0$}.} 
	\label{fig2}
\end{figure}

Figure \ref{fig2} displays numerical solutions for 
$b = 0.02$, $p = 0.25$, $\kappa = 1$, $v_0 = 0.2$, 
$\beta_{0} = 800$, $\beta_{1} = 0.20$, $\omega = 2\pi$, $\alpha = 40$, 
$\gamma = 100$, $\delta = 0.25$, {and $\xi=0$}. The blue, 
black, red, and green lines show the respective solutions for $s$, $e$, 
$i$, and $r$ variables. This result is for the interval of 10 years, 
discarding the first 10 years of transient. The solution is oscillatory 
with a period equal to $2\pi$. 
\section{Dynamical behaviour for $\xi = 0$}\label{sec_constant}
First, we study the effects of a constant vaccination rate on the dynamical 
system, i.e., $\xi = 0$ in Eq. (\ref{vaccinaperturbada}).

The influence of $v_0$ on the dynamical system behaviour is measured by 
the associated Lyapunov exponents computed for each pair $v_0 \times [\cdot]$, 
where $[\cdot]$ is one parameter from the model. In this work, we display 
the Lyapunov exponents in  colour scale. When the Lyapunov exponents are 
greater than zero, we plot the first largest one and when they are smaller than 
zero, we plot the second largest. We follow this procedure due to the fact that the 
transformation from non-autonomous to autonomous systems gives us a null 
exponent \cite{Brugnago2023}. We use the Wolf algorithm to compute the 
Lyapunov exponents \cite{Wolf1985} under initial conditions 
equal to $s_0 = 1 - e_0 - i_0$, $i_0 = 10^{-3}$, $e_0 = 0$. 
The Lyapunov exponents showed in this research are limited in the range 
$[-0.4:0.4]$ to obtain a standard colour code. Nonetheless, it is important 
mentioning that the exponents can be lower or higher than this range. 
The transient is fixed as $5 \times 10^5$ integration steps.

\begin{figure}[!ht]
	\centering
	\includegraphics[scale=0.75]{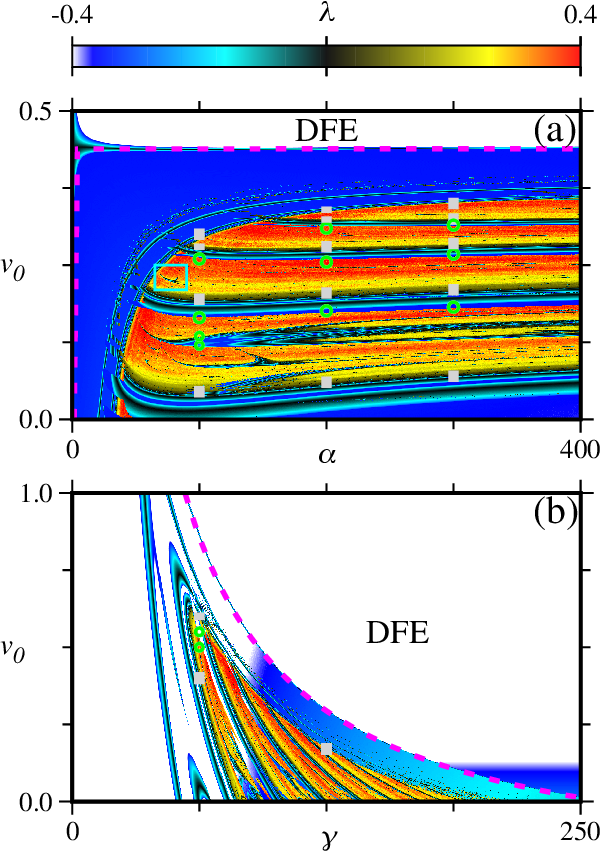}
	\caption{Influences of constant vaccination rate ($v_0$) and 
	$\alpha$, in the panel (a) for $\gamma = 100$; and $v_0 \times \gamma$, in the 
	panel (b) for $\alpha = 100$, in the Lyapunov exponent ($\lambda$), 
	in colour scale. 
	We consider 
	$b = 0.02$,
	$\omega=2\pi$,{}
	$\delta=0.25$,{}
	$\beta_{0}=270$,
	$\beta_1=0.28$, 
	and $p=0.25$. Green circles and grey squares highlight transitions by 
	crisis and period-doubling, respectively. The cyan square highlited in 
	the panel (a) is magnified in Fig. \ref{shrimp}. }
	\label{fig3}
\end{figure}

Figure \ref{fig3} exhibits the results for the parameter planes 
$v_0 \times \alpha$ and $v_0 \times \gamma$, in the panels (a) and (b), respectively.
The colour scale indicates the Lyapunov exponent ($\lambda$). 
DFE (disease-free) solutions are separated from non-DFE ones by the dotted magenta curve. To numerically 
verify the DFE, we compute the area under $i$ and $e$ curves. When both 
are less than $10^{-6}$, we obtain a DFE solution. 
Inspired by the minimal vaccination coverage obtained by Gabrick et al. \cite{Gabrick2024}, 
 we fit the numerical result for DFE using 
a function $v_0(\alpha) = a/(1 + b\alpha^{-1}) + c$, obtaining 
$a = 0.7089 \pm 0.0001$,
$c = 0.2702 \pm 0.0001$, 
where $b$ is the birth rate. 
Comparing with Eq. (13) from Ref. \cite{Gabrick2024}, we get  
$v_0(\alpha) = 0.715/(1 + 0.02\alpha^{-1}) - 0.27$. 
These results show that $\alpha$ has an influence on DFE only for small values.
For a fixed $\alpha$, variations of $v_0$ can lead the system from a periodic 
dynamics to chaotic via period-doubling, such as the points marked by the 
gray squares. 
Additionally, $v_0$ can lead the system  back 
to periodic behaviour via crisis \cite{Grebogi1983a} in some points, such as the points delimited 
by the green circles.  
{For a certain range of parameters, we observe the coexistence 
between periodic and chaotic attractors. 
When we compute the total number of infected individuals in the periodic 
or chaotic regime in the same time window, our results show that this 
number is practically the same for these parameter configurations. These results lead us to conclude that the 
total number of infected individuals is independent whether the dynamics are chaotic or periodic.}  

Fixing $\alpha=100$ and varying $\gamma$, we get the result shown in 
Fig. \ref{fig3}(b). 
The DFE separation is given by a non-linear expression equal to  
$v_0 (\gamma) = a/(\gamma + b) + c$, with 
$a = 70.47 \pm 0.02$,
$c = 0.2699 \pm 0.0002$, and $b$ is the natural birth. This relation establishes $v_0 \propto \gamma^{-1}$ 
as a threshold for DFE.  
The power-law behaviour can be understood by the fact 
that as the measure $\gamma$ increases faster the individuals go to $R$, which  
helps to obtain DFE once the other parameters are not able to sustain  
a endemic state.  
On the other hand, as $\gamma$ decreases until $\gamma = 56.56$, more 
vaccination is needed to obtain DFE. For values $\gamma<56.56$ even with 
$v_0 = 1.0$ there is no DFE. Contrary to the result in  panel (a), 
the type of dynamics is more stronger dependent on the combination of $v_0$ and 
$\gamma$. Nonetheless, for a fixed $\gamma$ value, period-doubling bifurcations  
and crisis also can be found, as marked by the gray squares and green circles, 
respectively. We also observe that the total number of infected does not change whether  
the dynamics is chaotic or periodic. 

\begin{figure}[!ht]
	\centering
	\includegraphics[scale=0.75]{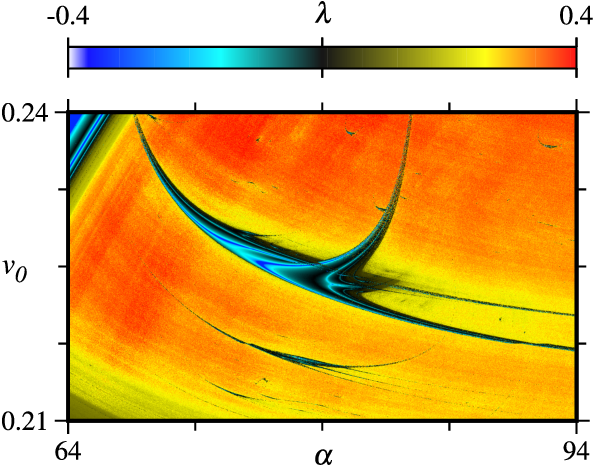}
	\caption{Magnification of a shrimp highlighted by the cyan square in 
	Fig. \ref{fig3}(a), for constant vaccination.  
	We consider: 
	$b  = 0.02$,
	$\omega=2\pi$,{}
	$\delta=0.25$,{}
	$\beta_{0}=270$,
	$\beta_1=0.28$, $\gamma=100$, 
	and $p=0.25$. }
	\label{shrimp}
\end{figure}

As previously observed, combinations of $v_0 \times \alpha$ lead to rich typical complex 
structures, such as shrimps,  highlighted by the small cyan square 
in Fig. \ref{fig3}(a). A magnification of this structure is display in Fig. \ref{shrimp}. 
Shrimps are periodic  structures immersed into  
chaotic bands \cite{Gallas1994,Martins2008}. 
As far we know, in epidemiological models the emergence of such structure 
was first reported in Ref. \cite{Brugnago2023}. Shrimps are 
important features because in their vicinity can happen cascades of similar periodic 
structures leading to a chaotic route \cite{Gallas1993b}. Close 
to shrimps, small changes in the parameters can drastically alter the dynamic \cite{Rech2017}.  
In our case, the main body of the structure has period 5. 

\begin{figure}[!ht]
	\centering
	\includegraphics[scale=0.75]{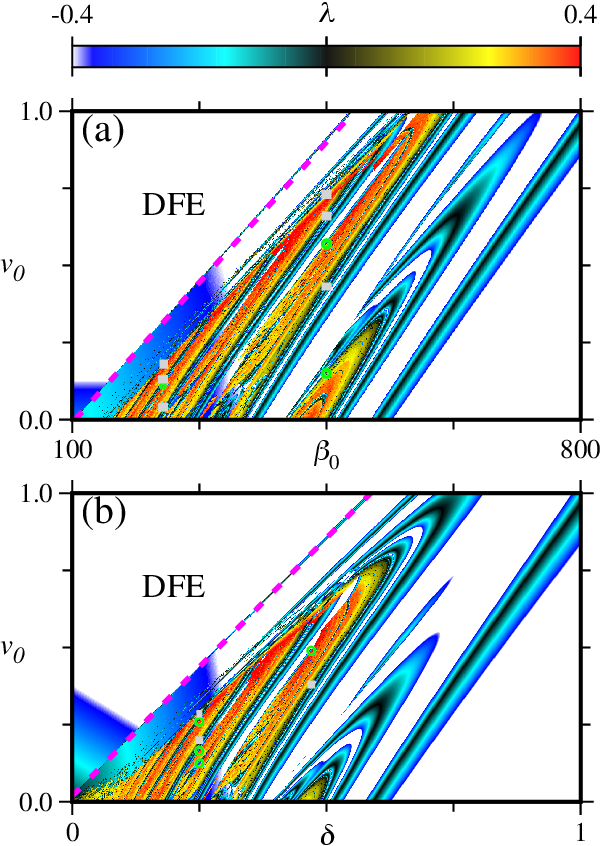}
	\caption{Influence of constant vaccination rate ($v_0$) and $\beta_0$, in the 
	panel (a) for $\delta=0.25$; and $v_0 \times \delta$, 
	in the panel (b) for $\beta_0=270$, in the Lyapunov exponent ($\lambda$), 
	in colour scale. 
	We consider 
	$b = 0.02$,
	$\omega=2\pi$,{}
	$\gamma=100$,{}
	$\alpha=100$,
	$\beta_1=0.28$, 
	and $p=0.25$. Green circles and grey squares highlight  
	crisis and period-doubling bifurcations, respectively. }
	\label{fig4}
\end{figure}

Figure \ref{fig4} exhibits the results for the combinations 
$v_0 \times \beta_0$ (panel (a)) 
and $v_0 \times \delta$ (panel (b)) as a function of $\lambda$ in the 
colour scale. These outcomes are for constant vaccination. 
For both panels, the DFE solution is delimited by a linear relationship 
given by $v_0 (x) = ax + c$. When $x=\beta_0$, we obtain 
$a=0.002 \pm 9 \times 10^{-7}$, 
$c=-0.2701 \pm 0.0003$, and for $x=\delta$: 
$a=1.6724 \pm 0.0004$ and 
$c=0.1848 \pm 0.0001$. The parameters are very close 
to the ones yielded directly from the theoretical result (Eq. (13) in Ref. \cite{Gabrick2024}). 
The apparent similarity 
of the results generated by varying $\beta_0$ and $\delta$ also occurs  
in the absence of vaccination \cite{Gabrick2023,Brugnago2023}.
As in the previous results, some bifurcations via period-doubling and crisis 
also are highlighted by grey squares and green circles. 
For $\beta_0 \geq 488.5$ only non-DFE solutions occur even for $v_0 > 0.95$ 
(Fig. \ref{fig4}(a)). Additionally, chaos is observed for $v_0 > 0.95$ 
in the range $\beta_0 \in (560, 614)$. This result shows that $v_0$ does 
not act as controling chaos. For  $\beta_0>604$ only periodic solutions 
remain, independent of $v_0$. In this case, a high contact rate leads to 
periodic solutions. 
Fixing $\beta_0 = 270$ and varying   
$\delta$, only periodic solutions exist for $\delta>0.624$ 
(Fig. \ref{fig4}(b)).   Therefore, 
when the lost of immunity occurs for periods less than 1.6 years, the 
dynamic is periodic. On the other hand, diseases with a lost of immunity 
greater than 1.6 can generate chaotic dynamics. 

\begin{figure}[!ht]
	\centering
	\includegraphics[scale=0.75]{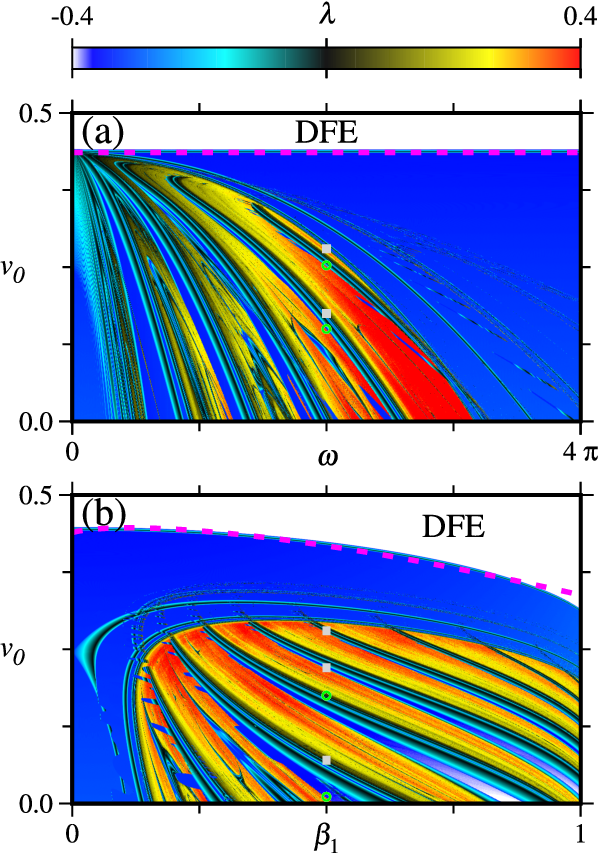}
	\caption{Effects of constant vaccination rate ($v_0$) and $\omega$, 
	in the panel (a) for $\beta_1=0.28$; and $v_0 \times \beta_1$, in the 
	panel (b) for $\omega=2\pi$, in the Lyapunov exponent ($\lambda$), in 
	colour scale. 
	We consider 
	$b = 0.02$,
	$\delta=0.25$,{}
	$\gamma=100$,{}
	$\alpha=100$,
	$\beta_0=270$, 
	and $p=0.25$. Green circles and grey squares highlight transitions by 
	crisis and period-doubling, respectively.}
	\label{fig5}
\end{figure}

Another constant relationship that separates DFE solutions from endemics ones 
is obtained by varying $\omega$ (Fig. \ref{fig5}(a)). In this case, the threshold 
is independent of $\omega$ and is given by $v_0 = 0.44$.  
For $v_0 < 0.44$ and $\omega \in (0.48, 3.26)\pi$
the solutions can be periodic or chaotic. Outside this range, there are 
only periodic solutions. For a given $\omega$ in the chaotic 
band, the increase of $v_0$ can suppress the chaotic behaviour. 
After some transient, the total infected only depends on $v_0$.
 One important value of this parameter is $2\pi$. 
We observe some transitions in the dynamics induced by crisis (green circles) 
and period-doubling (grey square). 

As our last analyses of the parameter plane, we fix 
$\omega = 2\pi$ and vary $\beta_1$, as shown in Fig. \ref{fig5}(b). 
For this result, the DFE is not described by the relation obtained in 
Ref. \cite{Gabrick2024}. This is expected once this minimum coverage is 
derived for non-autonomous model. Here, we give a contribution showing 
that for $\beta_1 > 0.5$ a non-linear contribution appear. Now the 
separation from DFE to endemic can be described by 
$v_0 (\beta_1) = a + c\beta_1 + d\beta_1^{e}$, with the parameters  
$a=0.4398 \pm 0.0003$, 
$c=0.7103 \pm 0.0006$, 
$d=-0.8114 \pm 0.0006$, and 
$e=1.108 \pm 0.001$. In this sense, an analytical expression for the minimum 
coverage for the forced model remains an open question. The non-linear 
contribution appears because for high levels of $\beta_1$   
combined with $v_0$ the  DFE is reach earlier. 
Chaotic solutions are found only for $v_0<0.38$ and $\beta_1>0.12$. 
Additionally, chaotic orbits depends are strongly dependent of $\beta_1$. 
Fixing $\beta_1 = 0.5$ and varying $v_0$, we observe many chaotic bands, 
that emerge from crisis or period-doubling bifurcation and are marked by green circles 
and grey squares, respectively. 

\section{Seasonal vaccination}\label{sec_seasonal}
Now, we consider the effects of a time-dependent vaccination campaign, 
i.e., $\xi$ and $\omega_{\rm v} \neq 0$ in Eq. (\ref{vaccinaperturbada}). 
However, to obtain the chaotic and bi-stable solutions we construct 
a bifurcation diagram for a constant vaccination rate by recording 
$i$ in the stroboscopic section as a function of $\beta_1$. 

\begin{figure}[!ht]
	\centering
	\includegraphics[scale=0.8]{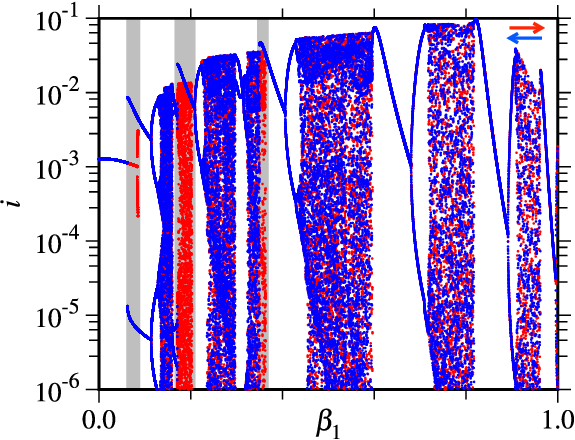}
	\caption{Bifurcation diagram type hystereses for $\beta_{1}$ considering 
	a constant vaccination rate $v_0 = 0.1$. The points are recording 
	in the stroboscopic section. 
	The red points 
	are in the forward and the blue in the backward direction of $\beta_{1}$. 
	The gray background delimits the bi-stable solutions.
	We consider $b = \mu = 0.02$, 
	$\alpha = 100$,
	$\gamma=100$,
	$\omega=2\pi$,
	$\delta=0.25$,
	$\beta_{0}=270$, and 
	$p=0.25$.} 
	\label{fig6}
\end{figure}

Figure \ref{fig6} displays the bifurcation diagram for $\beta_1$ 
and a constant vaccination rate $v_0 = 0.1$. The $y$-axis shows the $i$ 
variable in the stroboscopic section. We select these parametric configuration 
because the richness of the dynamical behaviour. 
The red and blue points are recorded in the forward and backward directions 
of $\beta_1$, showing a hysteresis curve \cite{Medeiros2017}. The grey background highlighted  
the bi-stable regions, which will be investigated in Section \ref{section_bistability}. 
In the absence of vaccination, for $\beta_1 > 0.9$ only chaotic solution 
is obtained  \cite{Gabrick2023,Brugnago2023}. 
For the considered parametric configuration and for $v_0=0.1$ a periodic 
solution emerges for high levels of $\beta_1$. It is worth mentioning 
that for other values of $v_0$ chaotic orbits can be found in this range 
(Fig. \ref{fig5}(b)). 
To explore the effects of $\xi$ and $\omega_{\rm v} \neq 0$ in the 
chaotic bands, we consider three levels of seasonality degree, which 
we define by: high $\beta_1 \in [0.7,1]$, 
medium $\beta_1 \in (0.3,0.7)$ and low $\beta_1 \in [0,0.3]$. In the 
next subsection we 
discuss in the following all of them.

\subsection{Low seasonality degree}
\begin{figure}[!ht]
	\centering
	\includegraphics[scale=0.8]{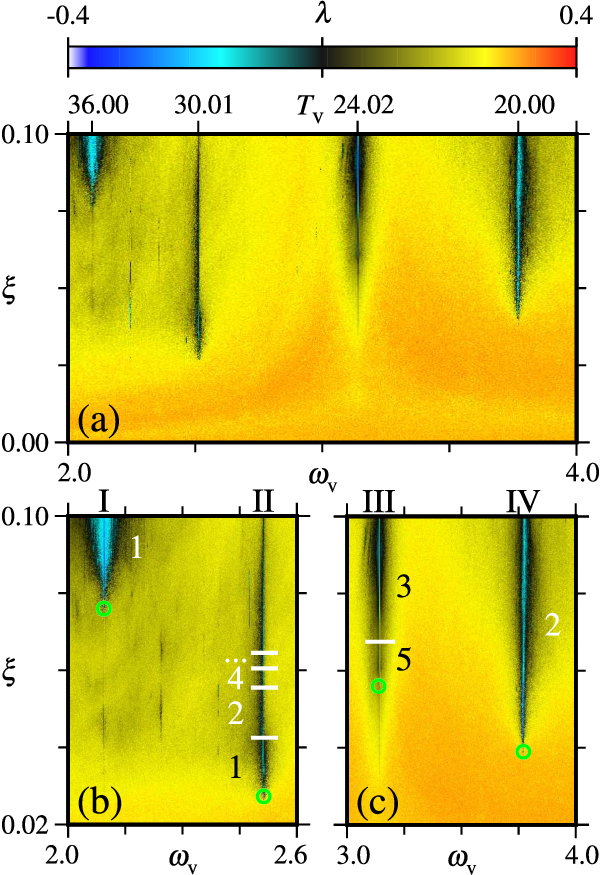}
	\caption{Impacts of seasonal vaccination in chaotic attractor for $\beta_1 = 0.15$. 
	Panel (a) display the parameter plane $\xi \times \omega_{\rm v}$ 
	where the colour scale indicates the Lyapunov exponents ($\lambda$). 
	The panels (b) and (c) exhibit the magnification of 
	periodic {bands} from the panel (a). 
	We consider $b = 0.02$, 
	$\alpha = 100$,
	$\gamma=100$,
	$\omega=2\pi$,
	$\delta=0.25$,
	$\beta_{0}=270$,
	$p=0.25$,
	and $v_0=0.1$.} 
	\label{fig7}
\end{figure}

Considering a seasonal vaccination campaign (Eq. (\ref{vaccinaperturbada}) 
with $\xi$ and $\omega_{\rm v} \neq 0$), we investigate how the chaotic 
solution for low seasonality ($\beta_1 = 0.15$) is affected.  
Figure \ref{fig7}(a) displays the parameter plane 
$\xi \times \omega_{\rm v}$ as a function of $\lambda$ in colour scale. 
It is important to note that in  Fig. \ref{fig7}(a), as well as in Fig. \ref{fig8}(a) 
and Fig. \ref{fig9},  the top $x$-axis exhibits the period $T_{\rm v}$ 
in months, where we mark the points in which  periodic {bands} emerge. 
In Fig. \ref{fig7}(a), we observe the existence of four periodic  {bands}, namely I, II, III, and IV. 
These {bands} occur for, and near, $\omega_{\rm v}$ ($T_{\rm v}$) =     
2.094 (36.00), 2.512 (30.01), 3.139 (24.02) and 3.771 (20.00). 
We select these points to analyse because they represent a  
richer dynamics than their neighbours. It is important to note that some 
periodic  {bands} are narrower than others. 
The first two  {bands}, I and II, are 
magnified in the panel (a). 

Fixing $T_{\rm v} = 36.00$, we increase $\xi$ 
from $0.02$ up to $0.10$, then a crisis is found at $\xi \approx 0.076$, 
which is marked by the green circle in Fig. \ref{fig7}(a). At this point, 
the chaotic attractors coalesce in a periodic branch with period 1.  
For which the amplitude of the $i$ variable in stroboscopic section decreases until 0.0056 in $\xi=0.1$. 

Following $\omega_{\rm v}$ in the positive direction, we reach the periodic 
{band} II (Fig. \ref{fig7}(a)). It 
displays  more complex dynamics when compared with I. For $T_{\rm v} = 30.01$ 
and increasing $\xi$ in the range from 0.02 to 0.1, a crisis occurs at  $\xi \approx 0.027$, 
and the chaotic attractor becomes periodic with period 1. 
The attractor with period 1 doubles the period at $\xi \approx 0.042$, which 
is denoted by a horizontal white line in  panel (a). 
The period 2 attractor suffers another period-doubling bifurcation at $\xi \approx 0.055$. 
The 4-period attractor ends in $\xi \approx 0.06$ and many subsequent 
period-doubling bifurcations occur, which we denote by the three white dots in the panel (b). 
For this value of $T_{\rm v}$, the chaotic attractor 
becomes periodic via a crisis (increasing $\xi$) and then returns back to chaos via period-doubling  
bifurcation. The return to chaos occurs in the range $\xi \in (0.064,0.1]$. 
The $i$ amplitude in the stroboscopic section is 
higher in the chaotic regime than in the periodic one. 

The {bands} III and IV are magnified in Fig. \ref{fig7}(c). 
To explore the dynamic associated with the {band} III, we consider 
$T_{\rm v} = 24.02$ and vary $\xi$. The chaotic 
attractor disappears via a crisis at $\xi \approx 0.056$ giving rise to a 
5-period attractor until $\xi \approx 0.067$. For $\xi > 0.067$ the 
period of the attractor changes to 3 and stays there for the 
whole analyzed range. We analyse the last {band}, namely IV, by 
considering $T_{\rm v} = 20$. A crisis occurs at  
$\xi \approx  0.039$ and the 
chaotic attractor changes to a periodic one with period equal to 2. This 
period does not change in the considered range. The {bands} III and IV leave 
the system near to DFE. In $\xi = 0.1$, the period 3 branch for III oscillates  
among $i = 0.0016$, 0.000009, and 0.006842 in the stroboscopic section. 
For the same value of $\xi$ and for IV the dynamic oscillates between 
$i = 0.006025$ and 0.000002. 

\subsection{Medium seasonality degree}
\begin{figure}[!ht]
	\centering
	\includegraphics[scale=0.8]{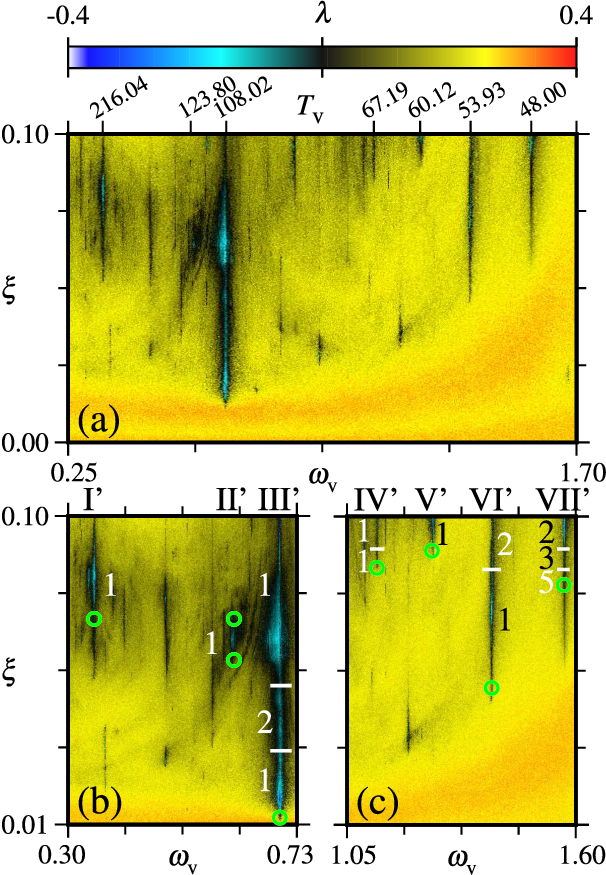}
	\caption{Impact of seasonal immunisation program in the chaotic attractor 
	for $\beta_1 = 0.5$. Panel (a) exhibit the parameter plane $\xi \times \omega_{\rm v}$ 
	as a function of Lyapunov exponent ($\lambda$), in colour scale.
	The periodic {bands} present in the panel (a) are magnified in 
	panels (b) and (c). 
	We consider $b = 0.02$, 
	$\alpha = 100$,
	$\gamma=100$,
	$\omega=2\pi$,
	$\delta=0.25$,
	$\beta_{0}=270$,
	$p=0.25$,
	and $v_0=0.1$.} 
	\label{fig8}
\end{figure}

To evaluate the seasonal vaccination campaign in a medium seasonality 
degree, we fix $\beta_1 = 0.5$. For this parameter, 7 periodic {bands} emerges   
in the range $\omega_{\rm v} \in [0.25, 1.70]$ 
($T_{\rm v} \in [301.59, 43.084]$), as shown in Fig. \ref{fig8}(a). 
The {bands} I', II', and III' are magnified in the panel (b), while 
IV', V', VI', and VII' in the panel (c). Each {band}  
occupies a certain range in the $\omega_{\rm v}$ ($T_{\rm v}$) parameter. 
Nonetheless, we select some values to locate each periodic branch. Focusing  
in the panel (a) and fixing $\omega_{\rm v} = 0.349$ ($T_{\rm v} = 216.04$), if we increase $\xi$ 
from 0 to 0.1 direction, we find a crisis point marked by the magenta 
circle that gives the origin for the {band} I'. The crisis occurs at  
$\xi \approx 0.07$ where the chaotic attractor coalesces in a periodic 
branch with period 1 and remains there until the analysed range ($\xi=0.1$). This periodic branch 
oscillates sinusoidal reaching a minimum value in the stroboscopic section 
equal to $i=0.016$. Thereafter, it starts increasing again. 

The periodic {band} II' can be obtaneid with $\omega_{\rm v} = 0.609$ 
($T_{\rm v} = 123.80$) and varying $\xi$ (Fig. \ref{fig8}(b)). 
In this case, the chaotic {band} goes to a 1-periodic 
attractor in $\xi \approx 0.058$ and remains there until $\xi = 0.07$. For values $\xi>0.07$ 
this {band} returns to the chaotic attractor. In the stroboscopic section, 
the periodic branch oscillates sinusoidal in $\xi \in (0.058, 0.07)$, 
having the first minimum local point in the pair $(\xi, i) \approx (0.061, 0.008)$. 
After that, it increases until a maximum in $(\xi, i) \approx (0.065, 0.037)$. 
From $\xi>0.065$ until $\xi=0.07$, the stroboscopic section in $i$ 
 decays and remains practically constant for $i \approx 0.0003$. 

To investigate the {band} III', we set $\omega_{\rm v} = 0.698$ 
($T_{\rm v} = 108.02$) (Fig. \ref{fig8}(b)). It occurs a crisis  
at $\xi \approx 0.01$ where the chaotic attractor go to 1-periodic 
attractor until $\xi \approx 0.031$. At this point, it occurs a bubble bifurcation  
 and the attractor starts to have period 2 until $\xi \approx 0.053$. 
After that, the attractor returns to period 1. The bifurcations are delimited 
in Fig. \ref{fig8}(b) by the white horizontal line. 

Now, we analyse  the range $\omega_{\rm v} \in [1.05, 1.6]$ 
($T_{\rm v} \in [71.807, 47.123]$), which is displayed in Fig. \ref{fig8}(c). 
The {bands} IV', V', VI', and VII' are narrow in 
$\omega_{\rm v}$ ($T_{\rm v}$)  when compared with I', II', and III'. 
Considering $\omega_{\rm v} = 1.122$ ($T_{\rm v} = 67.19$), we get the 
{band} IV', in which a 1-period attractor is created at $\xi \approx 0.085$ 
via a crisis. It is important to note that the 1-periodic attractor changes its 
amplitude for $\xi>0.0894$, which is delimited by the horizontal white line. 
Moreover, the attractor does not changes suddenly the amplitude. First, 
it occurs a period-doubling bifurcation in $\xi = 0.0885$ and  a new 
bifurcation in $\xi = 0.0887$ for a periodic attractor with period 3. 
Just after $\xi>0.0894$ that the attractor returns to period 1 with a different 
amplitude. We do not mark these bifurcations in the figure because it occupies a very 
short range in $\xi$. Following the 1-periodic attractor in the stroboscopic 
section for $i$, its amplitude increases linearly in the range $\xi \in [0.0894, 0.1]$, 
from $i \in [0.023, 0.027]$. 

A small periodic {band}, called V', also exists for $\omega_{\rm v} = 1.254$ 
($T_{\rm v} = 60.12$). 
The chaotic attractor goes to a 1-periodic attractor in the range $\xi \in (0.09,0.1]$. 
The structure VI' exhibits an interest dynamics for 
$\omega_{\rm v} = 1.398$ ($T_{\rm v} = 53.93$). The chaotic 
attractor goes to a periodic attractor via a bifurcation that starts in $\xi \approx 0.05$. 
In the range $\xi \in (0.05, 0.08)$, the periodic {band} has period 1 
and, at $\xi \approx 0.084$, a bifurcation occurs where the periodic attractor 
goes to period 2 and remains there. 

Finally, the last {band}, namely VII' (Fig. \ref{fig8}(c)), shows the transition from chaotic 
behaviour to periodic via a bifurcation at $\omega_{\rm v} = 1.571$ ($T_{\rm v} = 48$). 
Our results show that the chaotic attractors go to periodic ones via periodic doubling 
bifurcation. However, there is a periodic branch that occupies the range 
$\xi \in (0.06, 0.1]$, for $i = 0.0004$ in the stroboscopic 
section. The chaotic attractor becomes periodic with period 5 in $\xi \in (0.082, 0.083)$, 
where for $\xi \approx 0.083$ a new bifurcation occurs, that is marked in the 
panel (c) by the white horizontal line, and the periodic {band} becomes a 
3-period until $\xi \approx 0.091$. At this point, a new bifurcation occurs 
and the attractor has period 2. One branch for $i = 0.04$ and another 
for $i=0.0004$.

\subsection{High seasonality degree}
\begin{figure}[!ht]
	\centering
	\includegraphics[scale=0.8]{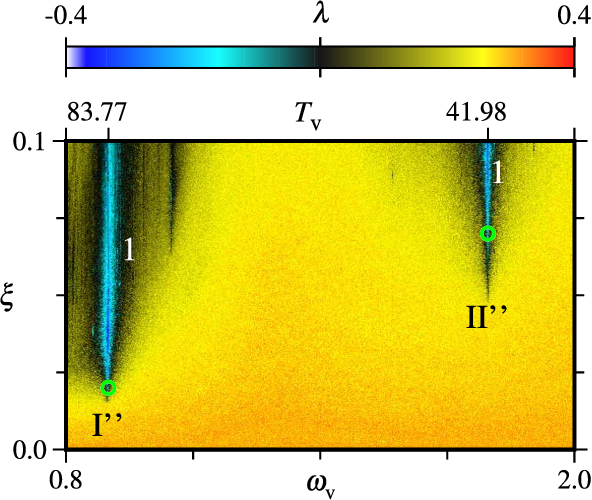}
	\caption{Impact of seasonal immunisation campaign in the chaotic attractor 
	for $\beta_1 = 0.95$.
	We consider $b = 0.02$, 
	$\alpha = 100$,
	$\gamma=100$,
	$\omega=2\pi$,
	$\delta=0.25$,
	$\beta_{0}=270$,
	$p=0.25$,
	and $v_0=0.1$.} 
	\label{fig9}
\end{figure}

Considering a high seasonality degree in the contact function, e.g., 
$\beta_1 = 0.95$, it is possible to obtain two periodic {bands} in the range 
$\omega_{\rm v} \in [0.8, 2]$ ($T_{\rm v} \in [94.247, 37.699]$), denoted by I'' and II'' 
(Fig. \ref{fig9}). 
The {band} I'' is wider (in $\omega_{\rm v}$ axis) and longer (in $\xi$ axis) 
than II''. 

Fixing $\omega_{\rm v} = 0.9$ ($T_{\rm v} = 83.77$), we observe that the 
chaotic attractor changes to a 1-periodic attractor at $\xi \approx 0.02$ (Fig. \ref{fig9}). 
The periodic branch increases in its amplitude. For example, if we consider 
$\xi = 0.01$ (chaotic regime), the maximum amplitude of the attractor is 0.01, 
in the stroboscopic section for $i$. 
However, when the attractor changes the regime, the periodic attractor increases   
in amplitude reaching the maximum value $(\xi, i) = (0.025, 0.062)$. 
For $\xi>0.025$, the $i$ amplitude decreases until $i=0.001$ in $\xi = 0.1$.

A different dynamics is generated by considering $\omega_{\rm v} = 1.796$ ($T_{\rm v} = 41.98$), 
which corresponds to  a value inside the {band} II'' (Fig. \ref{fig9}). The chaotic attractor becomes 
periodic in $\xi \approx 0.07$, but the amplitude of the attractor decreases, 
in the $i$ stroboscopic section from 0.07 (in chaotic regime) to 0.00004 (periodic behavior at $\xi=0.071$). 
Increasing $\xi$, the amplitude of the periodic branch also increases and reaches 
the pair $(\xi, i) = (0.1, 0.095)$. 

{In Figs. \ref{fig7}, \ref{fig8}, and \ref{fig9},  
some sparse points appear and the periodic bands are not well defined. 
By increasing the transient time or the number of iterations, we observe that 
the results remain unchanged. In this way, there is no transient effect 
present in the results. 
}
\section{Influence of seasonal vaccination in bi-stability}\label{section_bistability}
In this section, we investigate the effects of a time-dependent immunisation 
program in the basin of a bi-stable solution. This is made by using 
$\xi$ and $\omega_{\rm v} \neq 0$ (Eq. {\ref{vaccinaperturbada}}) and the 
basins are computed in the bi-stable solution showed in Fig. \ref{fig6}.  

Firstly, let us consider $\xi = \omega_{\rm v} = 0$. From the result in Fig. \ref{fig6}, we observe three bi-stable regions, 
that are highlighted by the gray backgrounds. In our simulations, we consider 
$\xi = \omega_{\rm v} = 0$ and $v_0 = 0.1$. These bi-stable regions shows 
the richness of the considered dynamical system \cite{Feudel1996a}. In 
this type of system the  evolution is extremely dependent on the initial 
conditions \cite{Feudel2008,Grebogi1983}. For 
more details about bi and multi-stable system we refer to Refs. \cite{Grebogi2020,Grebogi2017,Freitas2004,Menck2013}.

\begin{figure}[!ht]
	\centering
	\includegraphics[scale=0.7]{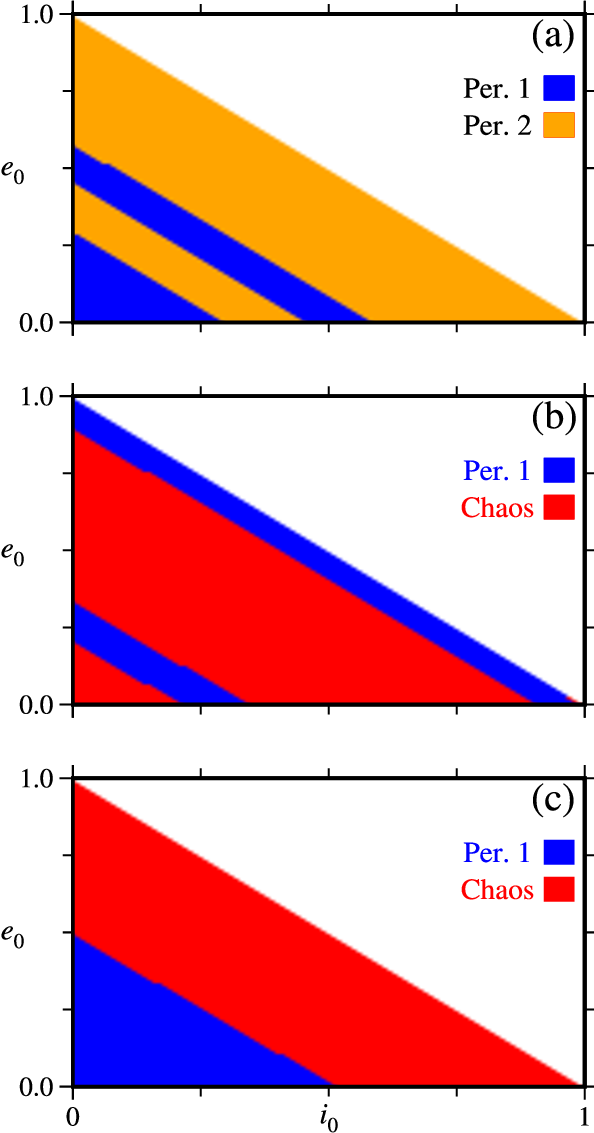}
	\caption{Basin of attraction for constant vaccination campaign ($\xi=0$).
	Each basin correspond to a section in Fig. \ref{fig6}, where 
    the panel (a) is for $\beta_1 = 0.07$, (b) is for $\beta_1 = 0.18$, 
	and (c) is for $\beta_1 = 0.36$. The blue colour shows points that 
	evolve to periodic attractor with period 1, orange exhibit initial 
	conditions that goes to period 2, and the red part is relative to 
	chaotic attractors. 
	We consider $b = 0.02$, 
	$\alpha = 100$,
	$\gamma=100$,
	$\omega=2\pi$,
	$\delta=0.25$,
	$\beta_{0}=270$,
	$p=0.25$,
	and $v_0=0.1$.} 
	\label{fig10}
\end{figure}

In Fig. \ref{fig6}, the first highlighted background shows the coexistence 
between two different periodic attractors in  $\beta_1 \in (0.06, 0.09)$.   
After $\beta_1 \in (0.06, 0.083)$, it occurs a bifurcation 
to period 2 in the red attractor. We compute the basin of attractions  
considering the pair $e_0 \in [0,1] \times i_0 \in [0,1]$, 
and $s_0 = 1 - e_0 - i_0$, such that $s_0 + e_0 + i_0 \leq 1$. We discretized our space in a grid of  
100 $\times$ 100. The basin for $\beta_1 = 0.07$ is displayed in Fig. \ref{fig10}(a). 
Only in this basin we do not follow the colour code from Fig. \ref{fig6}. 
Instead, we mark the basin correspondent to the period 2 attractor by the orange colour 
and for period 1 by blue one.  
 The white colour shows a prohibited region, 
where $s_0 + e_0 + i_0 + r_0 > 1$. Considering this mentioned discretization, we 
obtain 10000 possible initial conditions where 5050 evolve to one attractor 
or another. We call these 5050 initial conditions by valid points. 
We use the notation $\sigma$ 
for the fraction that evolves to a period 1, $\Delta$ for the 
fraction that goes to period 2, and $\Gamma$ for chaotic orbits. 
 The period of the orbits is determined from the stroboscopic 
section. From the valid initial condition, a fraction equal to 
$\sigma = 0.23$ evolves to the attractor with period 1, 
while the remaining $\Delta = 0.77$ evolves to period 2 attractor. 

The second bi-stable region is in $\beta_1 \in (0.17, 0.20)$ and exhibits   
the coexistence of a chaotic attractor, in the red branch, and a periodic one,  
in the blue branch. The basin for $\beta_1 = 0.18$ is displayed in Fig. \ref{fig10}(b). 
The red colour shows the pair $(e_0,i_0)$ that evolves to a chaotic attractor,  
while the blue colour displays the pair that goes to a periodic regime. Considering 
the same procedure, the fractions are $\Gamma = 0.75$ and 
$\sigma = 0.25$. For $\xi = \omega_{\rm v} = 0$, around 75\% 
of the considered initial conditions evolve to the chaotic regime.

The last bi-stable range is in $\beta_1 \in (0.35, 0.365)$ where there is  
a coexistence between chaotic and periodic attractors. Figure \ref{fig10}(c) 
displays the basin of attraction for $\beta_1 = 0.36$. In this case, 
$\sigma = 0.26$ and $\Gamma = 0.74$, i.e.,  the 
chaotic attractor is preferable. 
Besides we have attractor with period 1 in each basin, is important observe 
that they are different. 

Having the fraction of valid initial conditions that goes to period 1, 
2, and chaos under constant immunisation campaign, we explore how these 
fractions change in relation to $\xi$ and $\omega_{\rm v} \neq 0 $, 
i.e., a time-dependent campaign. To do this, we compute the basins by each pair  
$\xi \times \omega_{\rm v}$ and the fraction $\sigma$. Then, we verify how 
$\sigma$ depends on $\xi$ and $\omega_{\rm v}$ by looking at the parameter 
planes. We use the following colour code: $\sigma \in [0,0.05]$ in black; 
$\sigma \in (0.05,0.45]$ in yellow; $\sigma \in (0.45, 0.5]$ in white; 
$\sigma \in (0.5,0.55]$ in cyan; $\sigma \in (0.55, 0.95]$ in blue; 
and $\sigma \in (0.95,1]$ in purple. The intermediate colours shows the 
intermediates values.

\begin{figure}[!ht]
	\centering
	\includegraphics[scale=0.7]{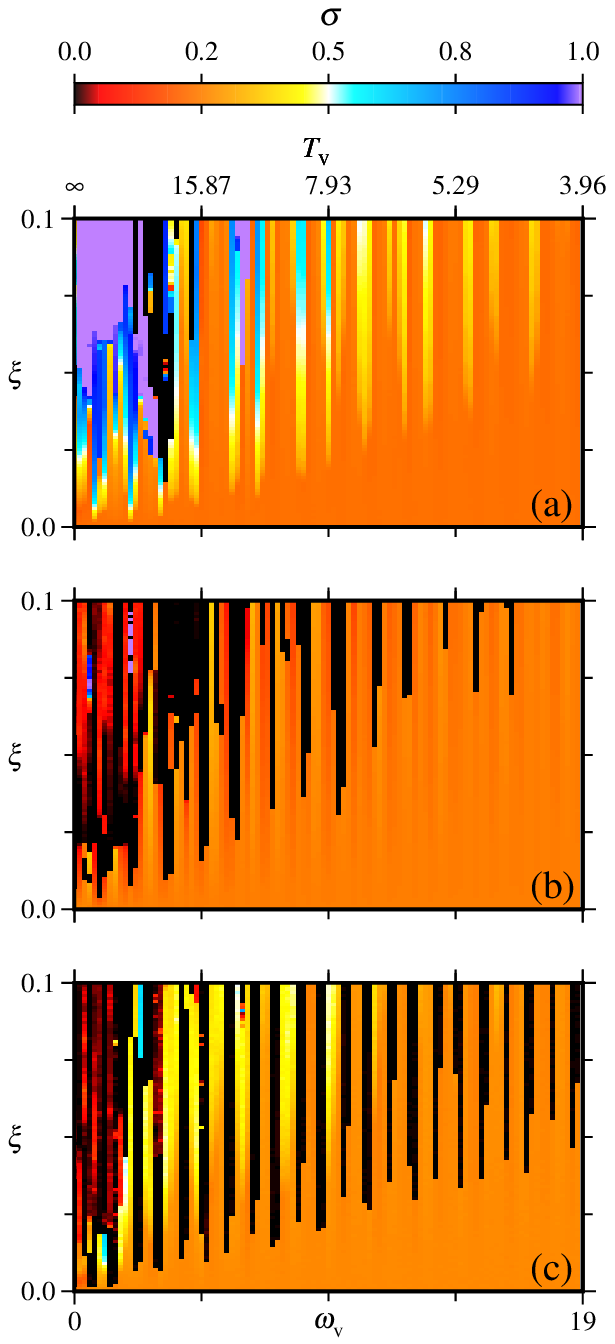}
	\caption{Parameter plane $\xi \times \omega_{\rm v}$. 
	{The colour scale displays the fraction of valid initial 
	conditions that evolve to period 1 attractor ($\sigma$).}
    The panel 
	(a) is for $\beta_1 = 0.07$, (b) is for $\beta_1 = 0.18$, 
	and (c) is for $\beta_1 = 0.36$.
	We consider $b = 0.02$, 
	$\alpha = 100$,
	$\gamma=100$,
	$\omega=2\pi$,
	$\delta=0.25$,
	$\beta_{0}=270$,
	$p=0.25$,
	and $v_0=0.1$.} 
	\label{fig11}
\end{figure}

Firstly, let us consider $\beta_1 = 0.07$, which corresponds  
to the basin in Fig. \ref{fig10}(a). 
Figure \ref{fig11}(a) displays the parameter 
plane $\omega_{\rm v} \times \xi$ where the colour scale represents $\sigma$. 
For a long period, i.e., $T_{\rm v} > 15.87$, it is possible 
to observe a significative purple range, i.e., more than 95\% of the valid 
points evolves to period 1 in this section. 
On the other hand, for 
$T_{\rm v} < 7.93$, the initial conditions evolve, mostly, to the 2-period 
attractor. It is worth to mentioning that mostly part of this parameter plane 
is marked by orange colour, which means that approximately 20\% of valid 
points goes to period 1.  

Considering $\beta_1 = 0.18$, there is a bi-stable dynamics where chaotic 
and 1-period attractor coexist. Emplying the same methodology that used 
to generate \ref{fig10}(a), we get Fig. \ref{fig11}(b). A significant 
part of the parameter plane is black, red, and orange. Meaning that 
$\sigma<0.5$ for mostly combinations of $\xi$ and $\omega_{\rm v}$. 
 As previously observed, without 
seasonal terms in the vaccination rate, almost  75\%  of the 5050 initial 
conditions evolve to the chaotic attractor. The preference by the chaotic 
attractor remains when we consider $\xi$ and $\omega_{\rm v} > 0$. 
Just small structures are in purple colour for $T_{\rm v}>15.87$. 

Another bi-stable solution exist for $\beta_1 = 0.36$. A similar result is displayed 
in the panel (c). In this case, we do not observe purple tones, only cyan. 
The rest of the parameter plane contain orange, black, and red colours, 
meaning that $\sigma<0.5$ for mostly part of $\xi \times \omega_{\rm v}$.

The methodology employed and the results shows that seasonal vaccination 
works as a control of 
bi-stability \cite{Feudel1997,Pisarchik2014,Goswami2008}.
\subsection{Why the chaotic attractors are preferred?}
Our results in Figs. \ref{fig11}(b) and \ref{fig11}(c) 
shows that reaching the chaotic attractor is
more preferable for the vaccination campaign. At this point, we need to understand 
the epidemics properties of periodic and chaotic attractors in an attempt 
to clarify why the chaotic attractors are preferred in this situations.   
As we observed in previous 
results, the total infected number is independent of chaotic or periodic attractors. However, 
{ for a given set of parameters, it is observed that chaotic 
attractors can have higher maxima values of infected individuals when compared 
with the periodic regime in the absence of vaccination \cite{Brugnago2023}. }
The same happen in presence of constant vaccination for some values of 
$\beta_1$ in the bi-stable range (Fig. \ref{fig6}). 
To understand  
which attractor is better in epidemic situations, we consider  two 
bi-stable regions in $\beta_1 = 0.18$ (Bi-stability 2) and $\beta_1 = 0.36$ 
(Bi-stability 3), as displayed in Fig. \ref{fig6}.

\begin{figure}[!ht]
	\centering
	\includegraphics[scale=0.7]{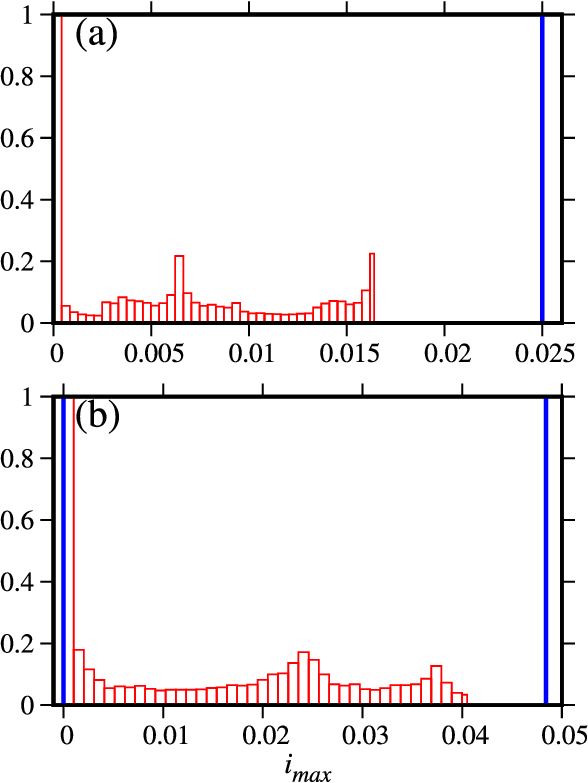}
	\caption{Distributions of maxima $i$ values for chaotic orbits (red bars) 
	and periodic (blue bars) under constant vaccination ($v_0 = 0.1$). The panel (a) is for  $\beta_1 = 0.18$ and 
	the panel (b) for $\beta_1 = 0.36$. 
	We consider $b =  0.02$, 
	$\alpha = 100$,
	$\gamma=100$,
	$\omega=2\pi$,
	$\delta=0.25$,
	$\beta_{0}=270$,
	$p=0.25$,
	and $v_0=0.1$.} 
	\label{fig12}
\end{figure}

One important characteristic in epidemic situations is the maximum point 
of infected individuals. In order to evaluate the local maximum points for 
the chaotic time series, we evolve the system during $2 \times 10^5$ 
iteration steps, discarding $10^{6}$ as transient, over all the chaotic initial 
conditions according to Figs. \ref{fig10}(b) and (c). In  
total, we have 3803 and 3730 initial conditions for $\beta_1 = 0.18$ 
and  $\beta_1 = 0.36$, respectively. Figure \ref{fig12}(a) shows the normalised 
frequency of maxima values of $i$ for $\beta_1 = 0.18$. The red bar is 
related to chaotic time series while the blue is periodic. We normalise 
according to the higher frequency value. From this result, we observe 
that the maxima point for periodic orbit is always in 0.025, while the 
chaotic are distributed in the interval $(10^{-6},0.017)$. The higher 
occurrence of maxima values of $i$ for the chaotic attractor occurs for 
$i_{\rm max} < 10^{-5}$. Figure \ref{fig12}(b) displays a similar result for 
$\beta_1 = 0.36$. The values of $i_{max}$ for the chaotic initial 
conditions are distributed in the interval $(10^{-6},0.041)$, with higher 
occurrence for $i_{\rm max} < 10^{-5}$. By computing $i_{\rm max}$ for the periodic 
initial conditions, we obtain a period 2, showing that the epidemic, in 
this case, is biannual, when we consider the section by the maxima points. 
In this case, we find $i_{\rm max} = 0.0484$ or  $i_{\rm max} = 2 \times 10^{-6}$. 
From both of these results, we observe that chaotic attractors have less maximum  
value than periodic ones. 

Another important measure is to know how long the attractor stays within limits 
of fewer and higher infections. As our model has the $E$ compartment, it is  
important to evaluate $i$ and $e$. Considering the same transient and the same 
initial conditions as in the previous case, we project the attractors in 
the plane $e \times i$ and establish thresholds in relation to $i$ and $e$ maxima 
values in the projection. We select the thresholds equal to 50\% and 
75\% of these pairs. By means of the thresholds, we 
compute how long the time series stays below ($t_{\rm B}$) and above 
($t_{\rm A}$) it. For the periodic case, the values are the same for 
every initial condition. For the chaotic situation, we see differences  
among the initial conditions, considering 3803 different 
initial conditions for $\beta_1 = 0.18$ (chaotic part of Fig. \ref{fig10}(b)) 
and 3730 for $\beta_1 = 0.36$ (chaotic part of Fig. \ref{fig10}(c)). 
Table \ref{tabela1} displays the results for $t_{\rm B}$ and $t_{\rm A}$ 
for the periodic and chaotic attractor, for $\beta_1 = 0.18$ and $\beta_1 = 0.36$, 
respectively. The times are calculated for a window time equal to 200 years. 
Nonetheless, we verify that $t_{\rm B} + t_{\rm A} < 200$. 
The relation $t_{\rm B} + t_{\rm A} = 200$ does not occurs because we are 
considering projections of the attractor. 
For both $\beta_1$ values, we observe that 
the chaotic attractor always spends more time below the threshold. For 
50\% and 75\% the chaotic attractor also spends less time in the upper part 
of the attractor projection in $e \times i$. {Combined with the information from Fig. \ref{fig12},  
we show that, under specific conditions, the chaotic attractor can present 
low levels of infection compared with periodic orbits in some situations.} Another important aspect when we are dealing with chaotic orbits 
is to know the horizon of predictability, which is $\propto 1/\lambda_1$. 
For $\beta_1 = 0.18$ and $\beta_1 = 0.36$, our simulations suggest 
$3.51 \pm 0.15$ and $3.02 \pm 0.13$, respectively. Our results indicate 
that proportionally to 3 years, we can predict our time series, after that, 
we need more simulations. In terms of epidemic prediction, this is a very 
reasonable forecast horizon.  
\begin{table}[!htb]
	\centering
			\caption{Time in which the periodic and chaotic attractor stay 
	below ($t_{\rm B}$) and above ($t_{\rm A}$) the thresholds  
	50\% and 75\% set from the projection of the attractor in the plane 
	$(i,e)$. For each chaotic attractor, we display the result $1/\lambda_1$ 
	computed from 3803 and 3730 chaotic initial conditions, for $\beta_1 = 0.18$ 
	and $\beta_1 = 0.36$, respectively. These results are for constant vaccination ($v_0 = 0.1$).}
	\begin{adjustbox}{width=0.47\textwidth} 
	\begin{tabular}{|c|cc|ccc|}
	\hline
						 && \multicolumn{1}{c}{$\beta_1 = 0.18$}  &&&\\
		\hline 			 &	 {Periodic}                      &&    Chaos          &&                          \\ \hline
		                 & $t_{\rm B}$    &  $t_{\rm A}$   &  $t_{\rm B}$         & $t_{\rm A}$    & $1/\lambda_1$           \\  \hline
		$50\%$           & 72.89          &   7.90         &  $147.28 \pm 1.99$   &    $6.55 \pm 0.28$  &  $3.51 \pm 0.15$   \\ \hline
		$75\%$           & 76.11          &   4.69         &  $151.78 \pm 2.20$   &    $2.43 \pm 0.15$  &  --                \\ \hline
		\hline
		\hline
						 && \multicolumn{1}{c}{$\beta_1 = 0.36$}  &&&\\
		\hline 			 &	 {Periodic}                      &&    Chaos          &&                          \\ \hline
		                 & $t_{\rm B}$    &  $t_{\rm A}$   &  $t_{\rm B}$         &      $t_{\rm A}$    & $1/\lambda_1$           \\  \hline
		$50\%$           & 47.10          &   3.75         &  $72.90 \pm 2.67$    &    $2.61 \pm 0.13$  &  $3.02 \pm 0.13$   \\ \hline
		$75\%$           & 48.75          &   2.10         &  $75.54 \pm 2.62$    &    $0.51 \pm 0.08$  &  --                \\ \hline
	\end{tabular}
	\end{adjustbox}
	\label{tabela1}
\end{table}  
\section{Conclusions}\label{sec_conclusion}
In this work, we study a SEIRS model with a periodic time-dependent 
transmission rate. In the first part of the manuscript, we consider a constant 
vaccination rate in the model. The vaccination is applied into the susceptible 
individuals at a rate $v_0$ and in the newborns at $p$. Then, we compute 
the Lyapunov exponents for the combinations of $v_0$ and another parameter 
from the model.  
We verify relationships between $v_0$ and another parameters of the model that lead the system 
to disease-free (DFE) solutions. In terms of dynamical behaviour, we show 
that the parameter planes exhibit a very rich dynamics, showing chaotic 
and periodic structures, in particular shrimps. We 
discover that even for high values of $v_0$ ($>0.95$), chaotic orbits 
exist.   

Besides the mentioned contributions, the main novelty of the paper was 
given by extending the vaccination program by including a time-dependent 
function. With this modification, we obtain very important results. 
Firstly, we discover that parameters related to the time-dependent vaccination 
can drive the chaotic attractors to a periodic one by generating periodic 
structures in the parameter plane. In mostly cases, these structures are 
acessed via a crisi or period-doubling bifurcation. 

{ It is important to mention that  we use the 
stroboscopic section to build the bifurcation diagrams and obtain the 
bi-stable solutions. Nonetheless, it is also possible to use the maxima infected number 
as a section. In this 
case, the bi-stability appears in the same range of the bifurcation diagram and  
some periodic attractors can have different periods. Furthermore, recording  
the maxima infected number as a function of the control parameter, the periodic attractor mostly  
has a higher amplitude in the infected individuals when compared to chaotic 
attractors in the bi-stable windows. }

By considering bi-stable solutions, between periodic-periodic and periodic-chaotic 
attractors, we propose a method to control these solutions based on the parameters 
of time-dependent immunisation program. 
This means that, with this methodology we 
can select one or other attractor as a function of vaccination parameters.  
We demonstrate that when chaotic and periodic 
attractors coexist, the basin of the periodic attractor is practically destroyed 
front a time-dependent immunisation campaign. This leads us to  a new result. Despite 
chaos seems a bad feature in epidemiological models due to the unpredictability, 
we observe the opposite. For a considerable number of initial conditions, 
we show that in certains sections of the bi-stable dynamics, the maximum point of infected 
individuals is lower for the chaotic attractor. In addition, given a threshold 
in the attractor projection in the plane $e \times i$, we find that chaotic attractor spend more 
time in the inferior part and less time in the superior part of projection 
than period orbits. Furthermore, typically, the inverse of the largest Lyapunov 
exponent is proportional to 3 years. 

{The  model discussed 
in this work shows very complex dynamics. One very characteristic important 
is the existence of a chaotic set in the multi-stable solution as an 
attractor and not as transient. Typically, in multi-stable systems 
chaotic attractors are difficult to be detected \cite{Feudel2003}. 
In certain cases, the chaotic dynamics appear as long transients. In our 
system, transient chaos also is possible for a given parameter configuration.  
It is important to note that as our model is an epidemiological model, long 
simulation times can be considered more than 100 years. For the bi-stable solutions discussed 
in this work, we consider transients until 1000 years and the chaotic 
attractor remains unchanged. A similar result was observed in 
the same model without vaccination \cite{Gabrick2023}. The inclusion of 
a new forcing increases the nonlinearity of the system, as a consequence 
we observe an increase in the chaotic basin. }

This works also adress some new problems. As far as we know, this is the 
second work that report the emergence of shrimps in epidemiological models 
(the first we refer to Ref. \cite{Brugnago2023}). In this way, questions 
like if this structure appear in other models or for some specific diseases 
remain open. Additionally, the structures reported in Fig. \ref{fig11} 
need to be investigate more deeply. 

\section*{Acknowledgements}
The authors thank the financial support from the Brazilian Federal Agencies (CNPq); 
S\~ao Paulo Research Foundation (FAPESP) under Grant Nos. 
2018/03211-6, 2020/04624-2, 2021/12232-0, 2022/13761-9, 2023/12863-5.
Coordenação de Aperfeiçoamento de Pessoal de Nível Superior (CAPES);  Funda\-\c c\~ao A\-rauc\'aria. 
E.C.G. received partial financial support from
Coordenação de Aperfeiçoamento de Pessoal de Nível Superior - Brasil (CAPES) - Finance Code 88881.846051/2023-01.
We thank 105 Group Science (www.105groupscience.com). 

\section*{DATA AVAILABILITY}
The data that support the findings of this study are available
from the corresponding author upon reasonable request.
\section*{References}
\bibliography{Ref}
\end{document}